# SARS-CoV originated from bats in 1998 and may still exist in humans

Ailin Tao[1,2,†], Yuyi Huang[2,3,†], Peilu Li[1,3], Jun Liu[3], Nanshan Zhong[2,*], Chiyu Zhang[1,3,*]

[1] Pathogen Diagnostic Center, Institut Pasteur of Shanghai, Shanghai Institutes for Biological Sciences, Chinese Academy of Sciences, Shanghai 200025, China

[2] Guangzhou Municipal Key Laboratory of Allergy & Clinical Immunology, the State key Laboratory of Respiratory Disease, the Second Affiliated Hospital of Guangzhou Medical University, Guangdong 510260, China

[3] Institute of Life Sciences, Jiangsu University, 301 Xuefu Road, Zhenjiang, Jiangsu 212013, China

SARS-CoV is believed to originate from civets and was thought to have been eliminated as a threat after the 2003 outbreak. Here, we show that human SARS-CoV (huSARS-CoV) originated directly from bats, rather than civets, by a cross-species jump in 1991, and formed a human-adapted strain in 1998. Since then huSARS-CoV has evolved further into highly virulent strains with genotype T and a 29-nt deletion mutation, and weakly virulent strains with genotype C but without the 29-nt deletion. The former can cause pneumonia in humans and could be the major causative pathogen of the SARS outbreak, whereas the latter might not cause pneumonia in humans, but evolved the ability to co-utilize civet ACE2 as an entry receptor, leading to interspecies transmission between humans and civets. Three crucial time points - 1991, for the cross-species jump from bats to humans; 1998, for the formation of the human-adapted SARS-CoV; and 2003, when there was an outbreak of SARS in humans - were found to associate with anomalously low annual precipitation and high temperatures in Guangdong. Anti-SARS-CoV sero-positivity was detected in 20% of all the samples tested from Guangzhou children who were born after 2005, suggesting that weakly virulent huSARS-CoVs might still exist in humans. These existing but undetected SARS-CoVs have a large potential to evolve into highly virulent strains when favorable climate conditions occur, highlighting a potential risk for the reemergence of SARS.

---

† These authors contributed equally to this study.

* Correspondence: C.Z., Pathogen Diagnostic Center, Institut Pasteur of Shanghai, Shanghai Institutes for Biological Sciences, Chinese Academy of Sciences, 411 Hefei Road, Shanghai 200025, China. Phone: 86-21-6385-0152, Fax: 86-21-6384-3571, E-mail: zhangcy1999@ips.ac.cn; or N.Z., the State key Laboratory of Respiratory Disease, Guangzhou Medical University, 250 Changgang Road East, Guangzhou, Guangdong 510260, China. Phone: 86-20-8306-2718, Fax: 86-20-8306-2729, E-mail: nanshan@vip.163.com.

## Introduction

Severe acute respiratory syndrome (SARS) coronavirus (SARS-CoV) is the causative pathogen of SARS and can be isolated from humans [1,2], palm civets [3,4], and bats [5,6]. Because SARS-CoVs from humans (huSARS-CoV) have a closer phylogenetic relationship with the strains isolated from civets (pcSARS-CoV) than those isolated from bats (batSARS-CoV), civets are naturally recognized as the immediate reservoir of huSARS-CoV [3,4]. A 29-nucleotide (nt) region, which can be found in the genomes of pcSARS-CoVs and early isolates of huSARS-CoVs but not in the genomes of the intermediate and later isolates of huSARS-CoVs, was used as compelling evidence to support the direct origin of huSARS-CoVs from civets [3,7,8]. However, another study demonstrated that the presence of the 29-nt region can not be used as evidence to trace the origin of huSARS-CoVs and argued against the cross-species transmission of SARS-CoV from civets to humans [9].

Two retrospective investigations revealed that in 2001 41.6% of 77 Beijing children and 1.8% of 938 Hong Kong adults were positive for SARS-CoV-specific antibody [10,11], whereas two other studies showed no seroprevalence of SARS-CoV among wild civets and other animals in Beijing and the southeastern provinces of China [4,12]. These data suggest that SARS-CoV existed among humans prior to the SARS outbreak and raise doubts about the direct origin of huSARS-CoV from civets. What the immediate reservoir of huSARS-CoV is, and when and why it originated, remain to be answered. Furthermore, since the SARS epidemic was well controlled, some experts believe that SARS-CoV has essentially disappeared as a threat. Recently, however, a new strain of coronavirus that causes acute serious respiratory illness similar to SARS appeared in the Middle East [13,14]. Although this virus has a closer phylogenic relationship with coronaviruses that infect bats in Southeast Asia than with SARS-CoVs [15], it raises a question as to whether SARS-CoV has been eradicated completely.

Here, we demonstrate that huSARS-CoV originated in 1998.6 through a cross-species jump from bats to humans and then further evolved into two lineages - one being highly virulent and associated with the SARS outbreak, and another being weakly virulent and then spreading from humans to civets. The occurrence of cross-species transmission of SARS-CoV and the SARS outbreak appear to be associated with anomalous climate changes in Guangdong. Furthermore, we find that SARS-CoVs may still exist among humans, sounding a global alarm on the possibility of the reemergence of SARS.

## Results

### Human SARS-CoV originated from bats rather than from palm civets

The spike (S) protein of SARS-CoV is responsible for viral attachment and entry into host cells and plays a crucial role in viral cross-species transmission [16,17]. To investigate the origin of SARS-CoV, we constructed a time-scaled maximum clade credibility (MCC) tree using SARS-CoV S gene sequences from humans, civets and bats for inferring the ancestral host state of viruses (Fig. 1). In the tree, all huSARS-CoVs and pcSARS-CoVs are divided into two sub-clades, one (sub-clade I) containing all 2003 huSARS-CoVs with one exception, and another (sub-clade II) including all pcSARS-CoVs and two huSARS-CoVs. Strains within sub-clades I and II have the most recent common ancestors (MRCAs) at nodes D and F, respectively, and share an earlier ancestor at node C. The ancestral host states of MRCAs were estimated to be exclusively human at nodes C, D and F (Fig. 1, left upper panel), indicating a transmission route from humans to civets [7,8]. All batSARS-CoVs form two sub-clades. One is located at the most basal position of the tree, and another clusters with the clade comprising huSARS-CoVs and pcSARS-CoVs and shares a MRCA at node B (Fig. 1). The ancestral host states of MRCAs most likely are bat at both nodes A and B (Fig. 1, left upper panel), suggesting a cross-species transmission from bats to humans.

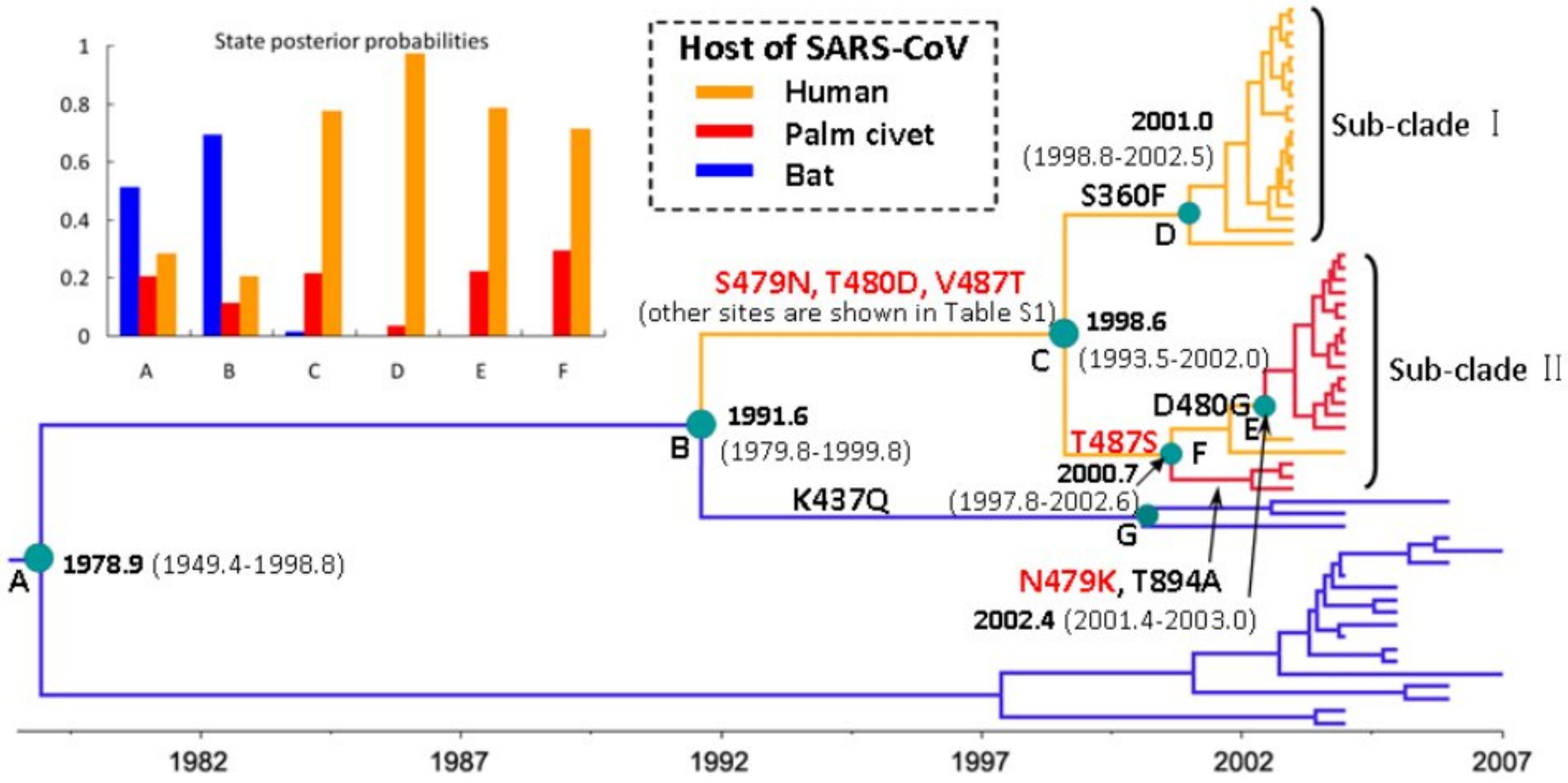

**Figure 1 Maximum clade credibility (MCC) tree of SARS-CoV based on spike genes.** Ancestral host states were reconstructed using Bayesian phylogeographic inference framework implemented in the BEAST v1.6.2 package. The host state posterior probabilities of the most recent common ancestors (MRCAs) are shown on the left upper panel. The tree branches are colored according to their respective host species. The solid nodes on the trees represent the MRCAs of corresponding virus strains. Estimated times of the MRCA (tMRCA) with 95% highest posterior density (HPD) are shown at corresponding nodes. The amino acid mutations between two ancestral S sequences at adjacent nodes are shown above the corresponding branches.

**Table 1. Amino acid mutations occurring in SARS-CoV S protein during evolution from node B to C.**

| Functional domain | Amino acid mutations | Numbers | Fisher's exact test (P) |
|---|---|---|---|
| Receptor-binding domain (RBD) | R322L, D326G, K327E, R333K, N336S, T343K, K344R, D347N, T353S, P371A, S372T, I375N, T380S, S381N, T386S, L388V, I389V, R390K, S391G, S392D, E393D, V397I, E401Q, T417M, I421L, A426R, K427N, Q428I, Q430A, G431T, K437N, Y439K, S442Y, H443L, K445H, T446G, K449R, L455I, S469P, D470P, E471A, G474C, V475Y, R476W, T477P, S479N, T480D, D482G, P485T, S486T, V487T, P488G, V489I, E490G, A493P, T494Y, | 56 | <0.0001 |
| Heptad repeat (HR) 2 | K1163E | 1 | 0.0415 |
| Other region | None | 0 | Not available |

The receptor-binding domain and heptad repeat (HR) 2 region of S protein contain 193 (residues: 318-510) and 44 (residues: 1142-1185) amino acids, respectively. The Fisher's exact test (Two-sided) was performed for RBD vs. other region and HR2 vs. other region. For other details, please see Figure 1.

RNA-dependent RNA polymerase (RdRp) is responsible for viral replication and determines viral survival [1,2]. A MCC tree based on the RdRp gene of SARS-CoV was also constructed (Fig. 2). Results similar to that of the S gene were observed. At node C leading to huSARS-CoVs and pcSARS-CoVs, the ancestral host state of MRCA was estimated to be human, confirming the transmission of SARS-CoV from humans to civets. At node B, the ancestral host state of MRCA was estimated to most likely be bat (Fig. 2, left upper panel), supporting the transmission route from bats to humans.

**Adaptive process of SARS-CoV from bats to humans**

Ancestral S protein sequences at nodes B, C, D, F and E were respectively reconstructed. Sequence comparison shows that there are 57 amino acid mutations occurring in the S protein from node B to C and one mutation occurring from node B to G (Fig. 1 and Table 1). The presence of only one mutation from node B to G indicates a long-term stable infection of SARS-CoV in bats, whereas the higher number of mutations seen from node B to C implies a complex adaptive process of SARS-CoV in humans after the bat-to-human cross-species jump. From node C to its downstream nodes, four amino acid mutations were identified. Of them, N479K, D480G, and T487S are involved in the use of civet angiotensin-converting enzyme-2 (ACE2) as the entry receptor for SARS-CoV [17], supporting the cross-species transmission from humans to civets (Fig. 1).

Ancestral RdRp protein sequences at the nodes B, C, and the downstream nodes were also reconstructed. Compared with that of the S protein, there are only three amino acid mutations occurring in SARS-CoV RdRp enzyme from node B to C, and one from node C to its downstream nodes (Figure 2). This suggests that the adaptation of SARS-CoV RdRp to the human host is easier than adaptation of the S protein.

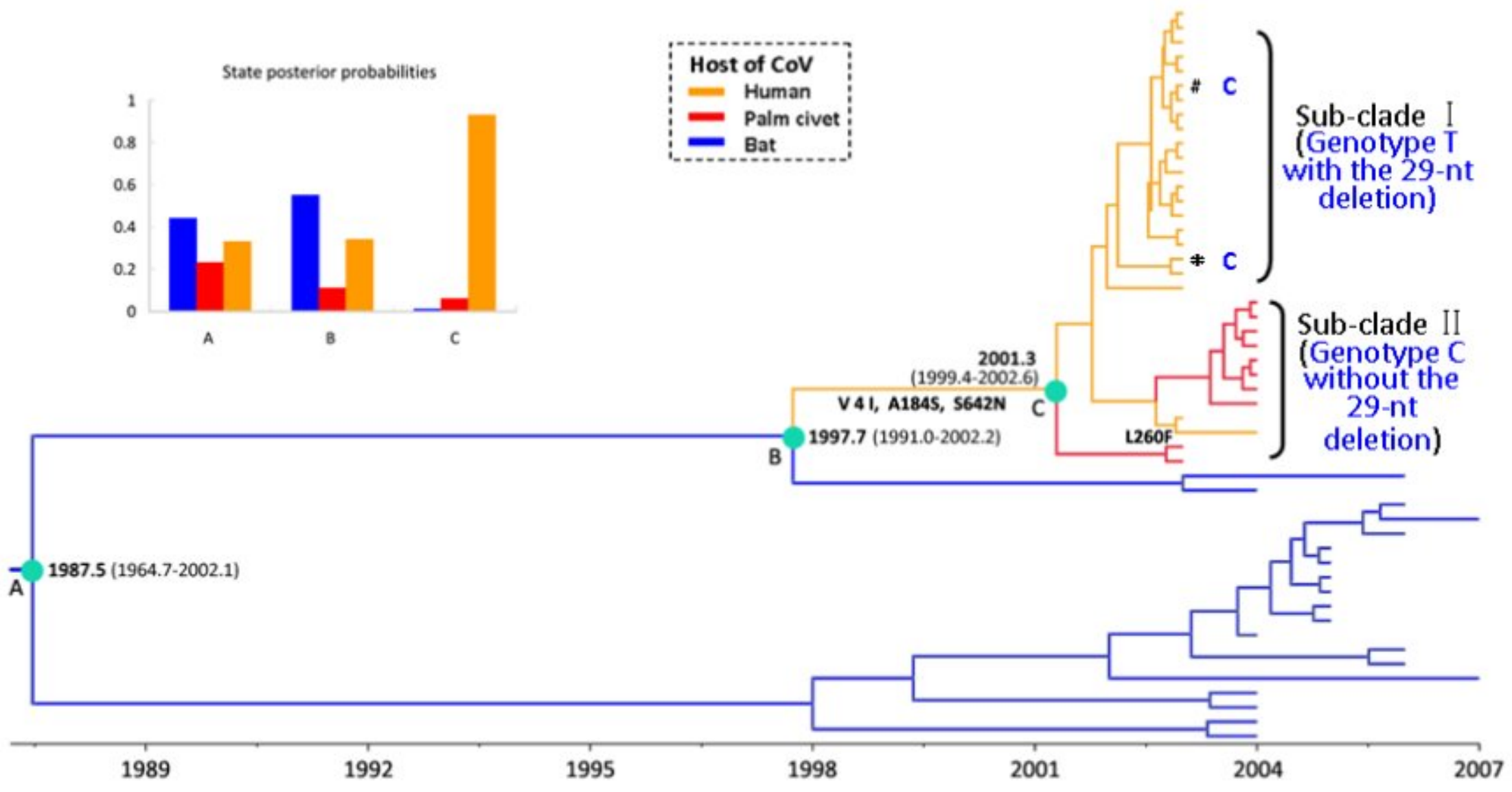

**Figure 2. Maximum clade credibility tree of SARS-CoV based on RdRp genes.** The symbol * indicates the strain with a genotype C characteristic of CCGAACGGCC and without the 29-nt deletion. The symbol # indicates the strain with a genotype C characteristic of CCGAACGGCC. For other details, please see Figure 1.

**SARS-CoV evolved into highly and weakly virulent lineages after crossing into humans.**

A previous study showed that SARS-CoVs could be classified into two genotypes, T and C, based on 10 polymorphism sites - 9404, 9479, 17564, 19838, 21721, 22222, 22517, 23823, 27243, and 27827, and the SARS-CoVs of the genotype T appeared to have a higher case fatality rate than those of the genotype C [18] . This suggests that the SARS-CoVs of genotype T may be more virulent than those of genotype C. By analyzing the genotypes of SARS-CoVs from two huSARS-CoV-associated sub-clades (I and II), we found that except for two strains at the basal position, all strains within sub-clade I belong to genotype T, suggesting their potential to be highly virulent, and all strains from sub-clade II belong to genotype C, suggesting they may be weakly virulent (Fig. 3). At the basal position of sub-clade I, one genotype C strain shares characteristics with those in sub-clade II, and another seems to be intermediate between genotypes C and T, implying an evolution from the weakly virulent genotype C strains to the highly virulent genotype T strains. This suggests that after jumping from bats into humans, SARS-CoV evolved into both highly and weakly virulent lineages.

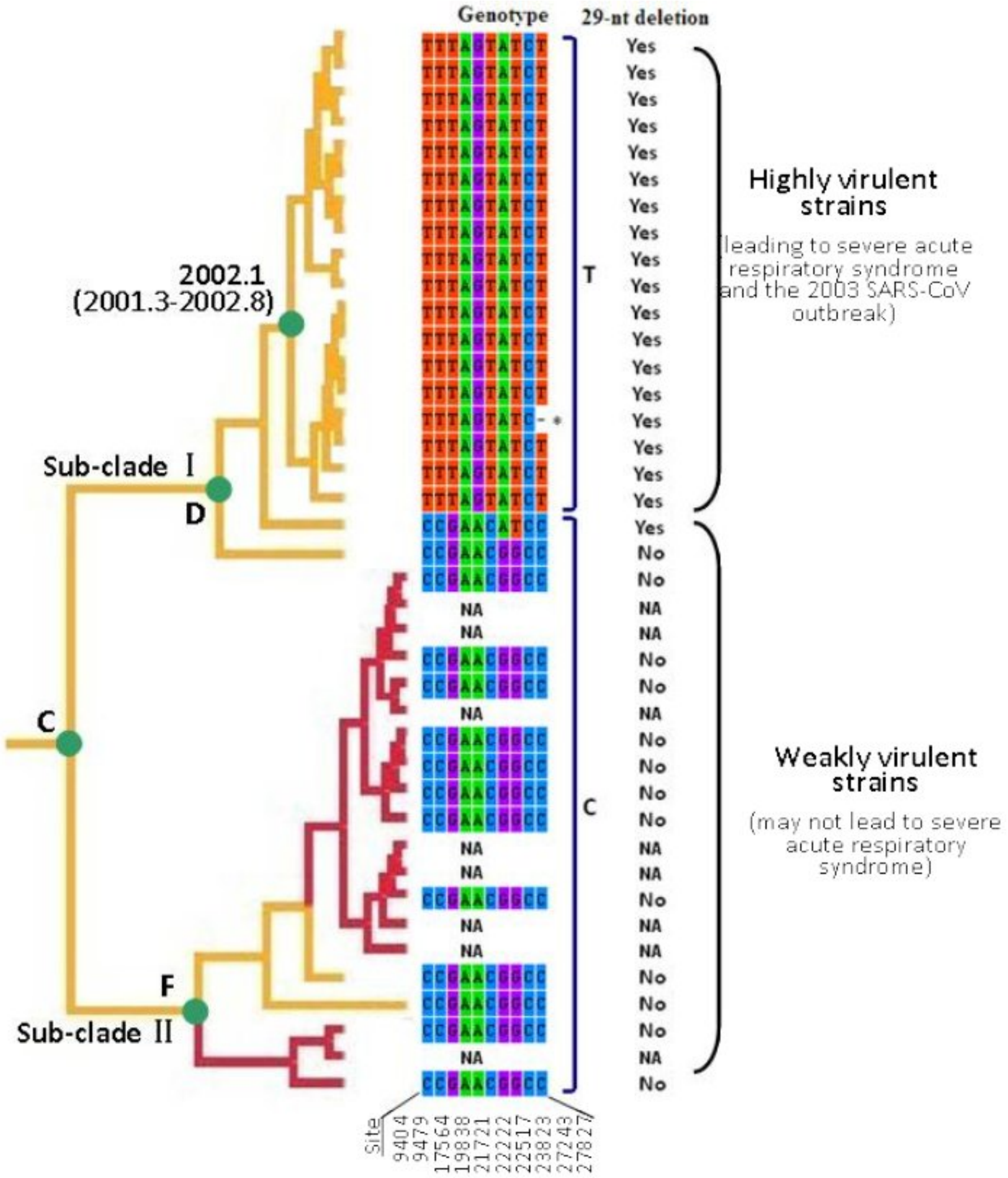

**Figure 3 The genotype characteristics of huSARS-CoVs and pcSARS-CoVs.** The sub-tree of the S proteins is obtained from Figure 1. The star indicates that the gap in site 27827 might be due to a sequencing error.

Some SARS-CoVs carried the 29-nt deletion, which was previously used as the predominant evidence for cross-species transmission of SARS-CoV from civets to humans [7,8]. We found that except for one strain at the basal position, all strains within sub-clade I carry the 29-nt deletion, whereas all strains from sub-clade II lack the 29-nt deletion (Fig. 3). Interestingly, all genotype T strains have the 29-nt deletion, and all genotype C strains appear to lack the 29-nt deletion except one genotype C strain at the basal position of sub-clade I that has the 29-nt deletion (Fig. 3). These indicate a close association between genotype T and the 29-nt deletion, suggesting that the 29-nt deletion may also be associated with the virulence of SARS-CoV.

On the other hand, because the MRCAs of both sub-clades I and II have the 29-nt sequence, the earliest human-adapted SARS-CoV strain (at node C in Fig. 1) should have the 29-nt sequence, and its ancestor batSARS-CoV is presumed to also have the 29-nt sequence. However, sequence comparison of the 29-nt region showed that huSARS-CoVs are more similar to the genetically distantly related batSARS-CoVs (batSARS-CoV sub-clade II) than their genetically closely related batSARS-CoVs (batSARS-CoV sub-clade I) (Fig. S1A and C), implying a recombination in this region. Bootscan analysis confirmed that the 29-nt region of huSARS-CoV originated by a recombination between the strains of batSARS-CoV I and II (Fig. S1B) [19].

### Temporal dynamics of SARS-CoV cross-species transmission

The time to the MRCAs (tMRCAs) at several crucial nodes in the MCC trees of both the S and RdRp genes was estimated. For the S gene, the tMRCAs at nodes B and C was estimated at 1991.6 (95%HPD, 1979.8-1999.8) and 1998.6 (95%HPD, 1993.5-2002.0), respectively (Fig. 1). This indicates that the SARS-CoV started to jump into humans from bats around 1991.6 and evolved into human-adapted huSARS-CoV strains around 1998.6. For the RdRp gene, the tMRCAs at corresponding nodes B and C was estimated at 1997.6 (95%HPD, 1991.0-2002.2) and 2001.3 (95%HPD, 1999.4-2002.6), respectively (Fig. 2), about 3-6 years later than the estimates based on the S gene.

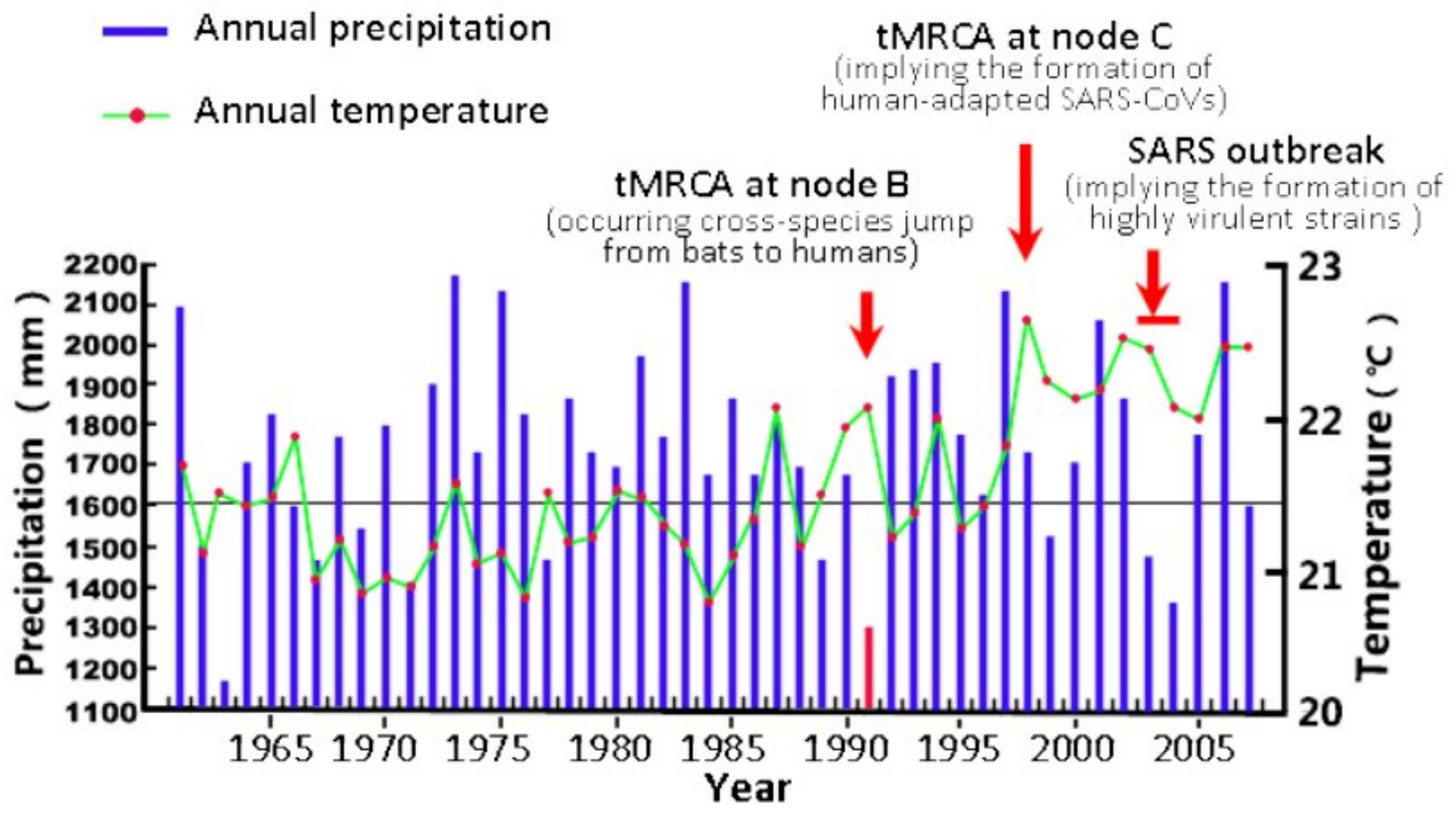

**Figure 4 Mean annual precipitations and temperatures at Guangdong, China from 1961-2007.** The data on annual precipitations and temperatures of Guangdong were obtained from the official reports available at http://www.tqyb.com.cn/NewTqyb/QHGB/GDGB2009.htm and http://www.djriver.cn/szygl/szygb.asp.

### An association between the cross-species transmission of SARS-CoV and climate conditions

Epidemiological studies [20,21] and phylodynamic analysis demonstrated that Guangdong is the geographic origin of huSARS-CoVs (Fig. S2). To investigate whether the cross-species transmission of SARS-CoV might be associated with regional climate conditions, we investigated the annual precipitation and temperatures of Guangdong from 1961 to 2007 (Fig. 4). We found that Guangdong experienced the second lowest annual precipitation and a relatively high annual temperature in 1991, which happens to correspond to the time when SARS-CoV started to jump from bats to humans (Fig. 4). Since 1991, Guangdong experienced two other times with very low annual precipitation and higher annual temperatures, in 2003 and 2004, well consistent with the time of the outbreak and reemergence of SARS-CoV. This type of climate condition also existed in 1998-1999, corresponding to the time of the appearance of the earliest human-adapted SARS-CoV (Fig. 4). These observations could suggest an association between the cross-species transmission of SARS-CoV and the unusual regional climate conditions.

**Table 2. Seroprevalence of SARS-CoV among population in Guangzhou before and after the 2003-2004 SARS outbreak.**

| Infection date | Sampling year | Number | Gender | | Age (years old) | | Anti-SARS-CoV positive number | Percentage (%) |
|---|---|---|---|---|---|---|---|---|
| | | | Male | Female | Mean | Scope | | |
| Infection before the SARS outbreak * | 2000-2001 | 8 | NS | NS | NS | NS | 6 | 75 |
| | 2002 | 28 | NS | NS | NS | NS | 17 | 60.7 |
| | Unknown but before 2002 | 6 | NS | NS | NS | NS | 5 | 83.3 |
| | Total | 42 | 19 | 15 | 22.3 | 3-42 | 28 | 66.7 |
| Infection after the SARS outbreak # | 2011 | 157 | 102 | 55 | 2.7 | ≤ 5 | 34 | 21.7 |
| | 2010 | 160 | 104 | 56 | 2.5 | ≤ 4 | 23 | 14.4 |
| | 2009 | 157 | 116 | 41 | 2.2 | ≤ 3 except one with 4 | 38 | 24.2 |
| | Total | 474 | 322 | 152 | 2.5 | ≤ 5 | 95 | 20.0 |
| Indeterminatio n | 2011 | 274 | 125 | 149 | 43.1 | ≥ 18 | 1 | 0.4 |

* The gender and age data are unavailable for 8 and 20 samples obtained before the SARS outbreak, respectively.
# All individuals were born after 2005, indicating that they were infected after the SARS outbreak.
NS: not shown.

**Presence of SARS-CoV before and after the 2003-2004 SARS-CoV outbreak**

We obtained serum samples that were collected from 42 individuals during 2001-2002, before the SARS outbreak, and detected IgG antibody against SARS-CoV. The result shows that 66.7% of these individuals were sero-positive (Table 2). We also determined the seroprevalence of SARS-CoV among Guangdong children who all were born after 2005. Among 474 sera samples obtained from these children, 95 (20%) tested positive for anti-SARS-CoV IgG antibodies (Table 2). In addition, during November and December of 2011, we collected blood samples from 19 previous SARS-patients and were able to detect IgG antibody against SARS-CoV in only two (10.5%) individuals (data not shown), indicating that the specific antibody against SARS-CoV can decrease and/or disappear over time.

## Discussion

Since the coronaviruses that are phylogenetically most closely related to huSARS-CoVs were isolated from palm civets, civets were indisputably considered to be the direct reservoir of huSARS-CoVs [3,4]. One reason is that animals often harbor a variety of viruses, so it’s logical to assume that animals are the natural reservoir of a new pathogen. Another reason is the discovery of the 29-nt deletion, which is strong evidence for the transmission of SARS-CoV from civets to humans, since it can be detected among all pcSARS-CoVs and early isolates of huSARS-CoVs, but not among middle and late isolates of huSARS-CoVs [3,7,8]. It was well known that civets played crucial role in expanding SARS-CoV transmission during the SARS outbreak [22,23]. However, if huSARS-CoV indeed originated from civets through cross-species transmission, we are unable to explain why SARS-CoV-specific antibodies were detected among children in Beijing and among adults in Guangdong and Hong Kong before the SARS outbreak (Table 2) [10,11], but not among wild civets and animals in Beijing as well as in 12 southeastern provinces of China [4,12]. Furthermore, Janies, et al clearly demonstrated that the 29-nt deletion does not support the transmission of SARS-CoV from civets to humans, and pointed out that the misleading conclusion drawn by previous studies was ascribed to the lack of an outgroup in previous phylogenetic analyses [9]. These observations imply that the direct origin of huSARS-CoV still remains elusive.

The BEAST phylogeographic analysis platform is a robust new tool that can be used to infer the geographic or host origin of a virus [24,25]. If the transmission route of SARS-CoV from civets to humans is correct, the host state of the MRCA of all huSARS-CoVs and pcSARS-CoVs should be the civet. The S and RdRp proteins are the two most crucial proteins for SARS-CoV survival [1,2]. The MCC trees of both the S and RdRp genes consistently show that the ancestral host of all huSARS-CoVs and pcSARS-CoVs is human, rather than civet. The Bayes factor (BF) test, which determines the statistically significant epidemiological linkage (BF>3), supports the transmission of SARS-CoV from humans to civets (BF = 23.73 and 120.38 for S and RdRp genes, respectively). These results, together with the detection of seroprevalence of SARS-CoVs in Beijing [10], Hong Kong [11] and Guangzhou in 2001 (Table 2), clearly indicates that the direct origin of huSARS-CoV was not from the civet, rather it was the reverse with pcSARS-CoV originating from humans. All batSARS-CoVs are located outside the clade comprising all huSARS-CoVs and pcSARS-CoVs, implying that the earliest huSARS-CoV most likely evolved from certain batSARS-CoVs. The BF test provides further support for the cross-species transmission of SARS-CoV from bats to humans (BF > 3.00 and = 3.38 for S and RdRp genes, respectively). In spite of this, we do not exclude the possibility that there might be intervening hosts involved in the movement of SARS-CoV from bats to humans.

To accomplish cross-species transmission, SARS-CoV must overcome at least two crucial bottlenecks [26,27]. The first one is that the S protein, which is responsible for viral attachment and entry into cells, must evolve the ability to recognize and use the receptor of different host cells [28]. After overcoming the receptor barriers, the second obstacle is that the replicase RdRp must adapt the intracellular environment of a new host to support productive replication of the viral genome [29]. Thus, the adaptation of the S protein to be able to recognize and bind to the new host receptor must be prior to the adaptation of the RdRp enzyme [28,29] and, as a result, the estimated time to achieve cross-species transmission of SARS-CoV based on the S gene should be sooner than that based on the RdRp gene. The date estimated for the cross-species jump of SARS-CoV from bats to humans based on the S gene is 1991.6, earlier than the estimated 1997.7 based on the RdRp gene. Regarding the tMRCA of huSARS-CoVs and pcSARS-CoVs, the estimates based on the S and RdRp genes are 1998.6 and 2001.3, respectively, supporting the above presumption that the adaptation of the S protein to human is prior to that of the RdRp enzyme.

Additionally, we note that the adaptation process of the S protein to a human receptor takes seven years, about two times longer than that needed for the RdRp enzyme. This indicates that the adaptation of the S protein is more difficult than the adaptation of the RdRp enzyme in order for SARS-CoV to cross-species jump from bats to humans, and implies that the adaptation of the S protein is the rate-limiting step in this process [28]. The ability of RdRp to effectively support replication of SARS-CoV in various cell lines confirmed that the adaptation of the RdRp enzyme is not the crucial step that limits the cross-species transmission of SARS-CoV [30]. Therefore, it was hypothesized that the S protein must undergo more mutations to overcome the species barrier than the RdRp enzyme does. As expected, comparisons of ancestral protein sequences show that the S protein accumulated 57 amino acid mutations during the adaptation process of SARS-CoV from bat hosts to human hosts, significantly more than the three amino acid mutations that occurred in the RdRp enzyme ($P<0.0001$, Fisher's exact test) (Table 1, and Fig. 2).

The receptor-binding domain (RBD) of the S protein is responsible for recognizing and binding to ACE2 [31], which is the functional receptor of SARS-CoV [16]. Heptad repeats (HR) 1 and 2 are involved in the membrane fusion function of the S protein [32]. Almost all (56, 98.2%) mutations occurring in the S protein were found to be located in the RBD, significantly more than those in HR region ($P<0.0001$) and other regions of the S protein ($P<0.0001$) (Table 1). In particular, these mutations in RBD include S479N, T480D, and V487T, all which were demonstrated to determine the receptor recognition and binding of SARS-CoV [17,33]. This strongly indicates that the most crucial step in the cross-species jump of SARS-CoV from bats to humans is the capacity of the S protein to evolve the ability to recognize and bind human ACE2, which has 258 amino acids that differ from bat ACE2 and indicates a substantial species barrier [34].

When a virus attempts to establish a stable infection in a new host, it often does not require too many mutations unless there is strong selective pressure from drugs or there is further cross-species transmission [28,35]. After the formation of human-adapted strains, SARS-CoV further evolved into two different lineages (sub-clades I and II) (Fig. 1). From node C to the sub-clade I strains, only one mutation was observed in the S protein, whereas from node C to the sub-clade II strains, a total of four mutations were observed. In particular, among these four mutations, N479K, S480G and T487S can all provide the S protein with the ability to co-utilize palm civet ACE2 for cellular entry of SARS-CoV [17,28,33]. These data strongly suggest that civets received the SARS-CoV infections from humans. After cross-species transmission from humans to civets, SARS-CoV could retain the ability to infect humans, which could have significantly contributed to the SARS outbreak by expanding the reservoirs for SARS-CoV transmission [22,23,36].

Genotypes T and C have been demonstrated to be associated with either high or weak virulence of SARS-CoV, respectively [18]. Among ten loci determining the SARS-CoV genotype, four (A21721G, C22222T, G22517A, G23823T) are located in the S gene, and none in the RdRp gene, implying that the S protein contributes to SARS-CoV virulence, but the RdRp enzyme does not. Among two huSARS-CoV-associated sub-clades in the S gene MCC, the sub-clade I comprises almost all huSARS-CoVs isolated in 2003, whereas the sub-clade II comprises two huSARS-CoVs and all pcSARS-CoVs isolated during 2003-2004. Except for two strains located at the basal position, all strains within the sub-clade I have genotype T, whereas all sub-clade II strains have genotype C (Fig. 3). In addition, all genotype T strains carry the 29-nt deletion, and, with one exception, all genotype C strains lack the 29-nt deletion (Fig. 3). This strongly suggests a close association between the genotype T and the 29-nt deletion, both of which together might determine the virulence of SARS-CoV. This information, together with the epidemiological observations that most 2003 huSARS-CoVs can result in pneumonia in humans [21,37] and even in civets [38], and whereas pcSARS-CoVs do not cause pneumonia in civets or other animals [3], indicates that the strains within sub-clade I are highly virulent and those within sub-clade II are weakly virulent [18]. Interestingly, within sub-clade I, the basal strain has a full genotype

C polymorphism map and lacks the 29-nt deletion, and the second basal strain has an intermediate polymorphism pattern between genotypes C and T, and carries the 29nt-deletion (Fig. 3). Therefore, the MRCA of all sub-clade I strains should belong to genotype C and have the 29-nt sequence. This implies that the highly virulent huSARS-CoVs leading to the SARS outbreak evolved from the weakly virulent strains (Fig. 3).

The time of formation of the highly virulent huSARS-CoVs was estimated to be 2002.1 (Fig. 3). This implies that during 1998.6 to 2002.1, the circulating huSARS-CoVs were weakly virulent strains that can cause asymptomatic self-limiting infections that resolve spontaneously. If so, the weakly virulent huSARS-CoVs may have been continuously circulating among humans since 1998.6 through to the present time. They may not be of concern because they cause non-pneumonic or mild infections. The serological data from two retrospective studies supported the presence of the weakly virulent huSARS-CoV strains before 2001 since there were no SARS-associated pneumonia cases reported before the SARS outbreak [10,11]. Furthermore, another study conducted in Hong Kong showed that even during the SARS outbreak the SARS-CoV-associated non-pneumonic infections were more common than SARS-CoV pneumonia [39], indicating that weakly virulent huSARS-CoV strains were co-circulating with other highly virulent strains during the SARS outbreak. However, because multiple effective control and intervention measures implemented in China and other countries helped stop the SARS epidemic quickly [40], some experts optimistically believe that SARS-CoV no longer exists in wildlife and has essentially disappeared as a threat.

To confirm the continuing presence of weakly virulent huSARS-CoV strains among humans, we performed serological tests on samples from 474 children who were all born in Guangzhou after 2005. Ninety five (20%) were found to be sero-positive for SARS-CoV among these children, but only one (0.4%) was sero-positive among 274 adults in Guangzhou. In addition, during 2000-2002, we found that 66.7% of inpatients (including children) in Guangdong were sero-positive for SARS-CoV (Table 2). The serological results most likely reflect the actual prevalence of SARS-CoV and are less likely to be due to cross reactivity of the ELISA kit with other human coronaviruses (huCoVs). Firstly, the microtiter plates (Beijing BGI-GBI Biotech Co., Ltd., Beijing, China) used to detect anti-SARS-CoV antibodies were coated with SARS-CoV lysate, which was demonstrated to have high sensitivity and specificity for SARS-CoV antibody testing [41]. Secondly, to reduce and/or exclude the possibility of cross reactivity, we randomly selected 271 and 90 samples from 474 children and 274 adults, respectively, in order to test for IgG antibodies against other huCoVs (Table S1). Among 132 child samples known to be anti-huCoV antibody negative, 26 samples tested positive for SARS-CoV-specific antibodies. Of particular importance is that among 36 anti-huCoV antibody positive adult samples, none was positive for SARS-CoV-specific antibodies. These results demonstrate that the SARS-CoV lysate ELISA IgG Kit that we used is highly specific for SARS-CoV antibody.

The serological results suggest that huSARS-CoV continues to exist after the 2003 SARS outbreak. Furthermore, these results, together with the observation that approximately 40% of Beijing children were sero-positive in 2001 at a time when only 1.8% of Hong Kong adults were sero-positive, might suggest that the weakly virulent huSARS-CoV preferentially infected children [10,11,40]. Because Guangdong is the geographic origin of huSARS-CoV (Fig. S2)[21], the very high prevalence (66.7%) of SARS-CoV in Guangdong during 2000-2002 might reflect a critical turning point resulting in the SARS outbreak.

Why did SARS-CoVs jump from bats to humans in 1991, and why did the SARS outbreak occur in 2003? We found that three crucial time points: 1991, when there was a cross-species transmission of SARS-CoV from bats to humans; 1998, when human-adapted SARS-CoV emerged; and 2003, when the SARS outbreak occurred, correspond well with the lower annual precipitation and relatively higher annual temperature observed in Guangdong during those years (Fig. 4). In particular, the cross-species transmission of SARS-CoV occurred in 1991, a year with the lowest annual precipitation levels (an exceptionally dry period) seen in Guangdong in the past three decades. Previous studies showed that low absolute humidity favors the airborne survival and transmission of respiratory viruses (e.g. influenza virus) [42,43,44]. Because the data on absolute humidity in Guangdong in the past three decades are not available, we were unable to determine whether change of absolute humidity in Guangdong was associated with the cross-species transmission and outbreak of SARS-CoV. Therefore, whether or not the exceptionally dry climate in Guangdong in 1991 facilitated the airborne survival of batSARS-CoVs and its cross-species transmission from bats to humans deserves further investigation.

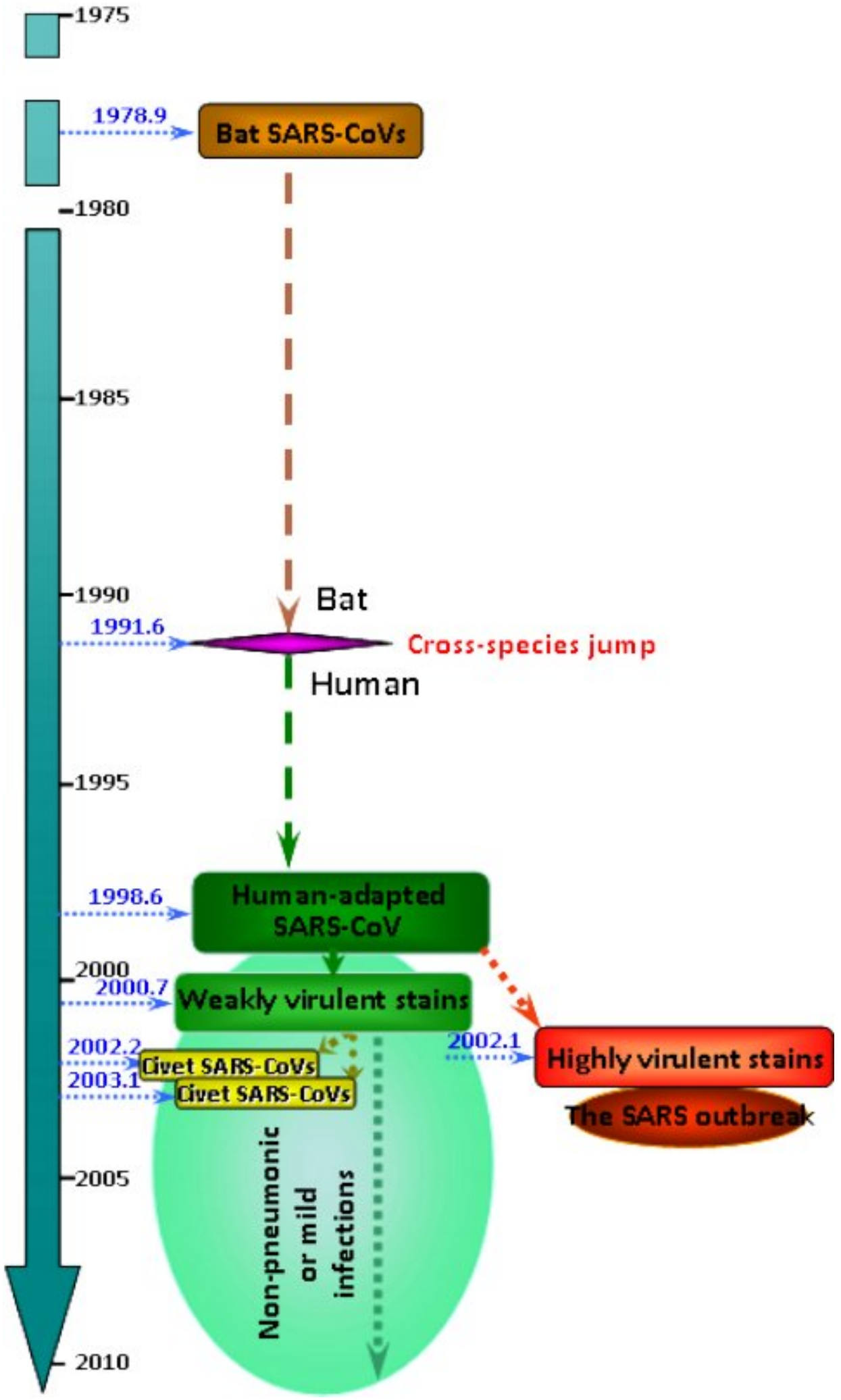

**Figure.5 Evolutionary history of SARS-CoV.**

In sum, we show that the earliest ancestor of SARS-CoV appeared in bats in 1978.9 (95%HPD, 1949.4-1998.8) (Fig. 5). In 1991.6, SARS-CoV began its cross-species jump from bats to humans and formed a human-adapted strain in 1998.6. After moving into human hosts, huSARS-CoVs diverged into highly virulent strains with genotype T and a 29-nt deletion mutation in 2002.1, and weakly virulent strains with genotype C but without the 29-nt deletion in 2000.7 (Fig. 5). The former strains can cause pneumonia in humans and could be the major causative pathogen of the 2003 SARS outbreak, whereas the latter strains might not cause pneumonia in humans but evolved the ability to co-utilize civet ACE2 as its entry receptor, leading to two waves of interspecies transmission between humans and civets (Fig. 5). Currently, the weakly virulent huSARS-CoVs may still exist among humans. Of concern is the possibility that they could evolve into highly virulent strains when favorable climate conditions occur, highlighting a potential risk of the reemergence of SARS [45]. Therefore, routine surveillance for the detection of huSARS-CoVs together with increased attention to climate changes are strongly urged to prevent the reemergence of SARS.

## Materials and Methods

### Ethics Statement

This study was done according to the Helsinki II Declaration and was approved by the medical ethics committee of the Second Affiliated Hospital of Guangzhou Medical University. To investigate the seroprevalence of SARS-CoV, we collected sera or plasma from 274 adults and 474 children during 2009-2011 and stored them at -80°C in Guangzhou Municipal Key Laboratory of Allergy & Clinical Immunology, the Second Affiliated Hospital of Guangzhou Medical University, until analysis. Verbal informed consents from the adult participants and the parents of the child participants for allergen diagnosis and other serological tests were obtained. Because this study was not involved in any individual privacy, the institutional review board waived the need for written informed consent from the participants. In addition, we also collected plasma samples from 19 previous SARS patients with written informed consents in December 2011.

### Sequence data collection

A total of 139 Spike (S) gene sequences of SARS-CoVs were retrieved from the GenBank according to three previous studies [3,5,7,35]. Among them, 59 S gene sequences have known times of sample collection, including 19 from bats, 18 from palm civets and 22 from humans, and were used to construct a time-scaled maximum clade credibility (MCC) tree for inferring the ancestral host sources of SARS-CoV [25,46]. The MCC tree shows a topology consistent with a previous study [28]. In addition, 102 S gene sequences from humans were used to infer the geographic origin of SARS-CoV.

### Reconstruction of time scale and ancestral host sources of SARS-CoVs

The codon alignment of all sequences was performed using the CLUSTAL W program implemented in MEGA 5.05 [47]. The MCC trees were constructed with a Markov Chain Monte Carlo (MCMC) method implemented in the BEAST v1.6.2 package, which combines time-of-sampling information and generates rooted trees [24]. The initial 25% of the trees were discarded as burn-in when we summarized the trees using TreeAnnotator v1.6.2. All of these trees were edited and viewed with FigTree v1.3.1. The evolutionary rates and the times to the most recent common ancestors (tMRCA) of various nodes in the MCC tree were estimated by BEAST. A relaxed molecular clock with an uncorrelated log-normal distribution and a constant population size model were used in the MCMC analyses. The codon-based substitution model (SRD06) was chosen as the nucleotide substitution model. The 95% highest probability density (HPD) was used to reflect the statistical uncertainty in parameter estimates. MCMC analysis was run for at least 1000 million generations, with sampling every 10,000 generations. The program Tracer v1.5 (tree.bio.ed.ac.uk/software/tracer/) was used to check for the convergence and to determine whether effective sample size (ESS)>200.

Each sequence was assigned a character state reflecting its host source or sampling location. Ancestral host source or location state was inferred using a geographically explicit Bayesian MCMC method under the asymmetric CTMC model for discrete state reconstructions implemented in the BEAST v1.6.2 package [25,46]. The ancestral host source and geographic origin were accessed by posterior probability that is calculated based on the posterior density of trees.

**Ancestral sequence reconstruction**

To investigate the adaptive changes in the S gene during SARS-CoV cross-species transmission, the ancestral S protein sequences at several crucial nodes were reconstructed using MEGA 5.05 package with maximum-likelihood (ML) method based on the topology of MCC tree [47]. The best-fitting nucleotide substitution model of TN (Tamura-Nei) 93 was selected.

**Seroprevalence of SARS-CoVs among children in Guangdong**

To determine whether SARS-CoVs are still circulating among humans, we investigated the current seroprevalence of SARS-CoV in children in Guangdong, China. Stored sera from 474 children were obtained from the Second Affiliated Hospital of Guangzhou Medical University. The samples were collected during 2009-2011 when these children had acute tracheobronchitis, and included 157 samples from 2009, 160 samples from 2010 and 157 samples from 2011. All children were born after 2005 with average age of 2.5 years old at the time of sample collection (Table 2). In addition, 274 stored adult sera were also obtained from the First and Second Affiliated Hospital of Guangzhou Medical University. To determine whether SARS-CoVs were circulating among the population in Guangdong before the 2003-2004 SARS outbreak, stored sera from 42 individuals who received bone marrow transplantation between 2000 and 2002 were obtained from the Guangzhou Blood Center. These sera were collected before the SARS outbreak.

Because the IgG antibody against SARS-CoV has a longer half-life in serum than the IgM antibody [48], the SARS-CoV lysate ELISA IgG Kit (Beijing BGI-GBI Biotech Co., Ltd., Beijing, China) was selected to specifically detect IgG antibody against SARS-CoV. This kit has high sensitivity and specificity for SARS-CoV (Ref) and was approved by the State Food and Drug Administration (SFDA) for the detection of anti-SARS-CoV immunoglobulin (Ig) G antibody from human serum or plasma specimens. To reduce and/or exclude the possibility of cross reactivity with other huCoVs, tests for other huCoV infections were also performed using a huCoV ELISA IgG Kit (Shanghai Lianshuo Bio-Tech Co., Ltd, Shanghai, China. The reagents (including antigens and antibodies) used in this kit were imported from Rapidbio, an US reagent company. All assays were performed according to the manufacturer's instructions.

**Acknowledgements**

We would like to thank Yinle Qi, Hui-Fang Chen, Ze-Hong Zou, Xue-Ting Liu, Jun-Yan Zhang, Ying He, He Lai, Nuo-Fu Zhang and Xiao-Guang Ye for their assistance in experiments. We also thank Lucinda Beck for her editing and critical reading of the manuscript and Dr. Ji-Fu Wei for his valuable suggestions for the manuscript. This work was supported by the National Natural Science Foundation of China (81071391) and the Natural Science Foundation of Jiangsu Province China (BK2011474), the National Key Science & Technology Special Projects on Major Infectious Diseases (2012ZX10004-211), and in part by Great Project from the Major Program of National Science and Technology of China (2011ZX08011-005) and Guangdong Provincial Natural Science Foundation (8251018201000002).

# Supporting Information

**Table S1. Detection of SARS-CoV- and huCoV-specific antibodies.**

<table>
<tr><td colspan="3" rowspan="2"></td><td colspan="4">Detection of SARS-CoV-specific antibodies</td></tr>
<tr><td>Positive</td><td>Negative</td><td>Total</td><td>Percentage (%)</td></tr>
<tr><td rowspan="8">Detection of huCoV-specific antibodies</td><td rowspan="4">Child samples</td><td>Positive</td><td>39</td><td>100</td><td>139</td><td>51.3</td></tr>
<tr><td>Negative</td><td>26</td><td>106</td><td>132</td><td>48.7</td></tr>
<tr><td>Total</td><td>65</td><td>206</td><td>271</td><td></td></tr>
<tr><td>Percentage (%)</td><td>24.0</td><td>76.0</td><td></td><td></td></tr>
<tr><td rowspan="4">Adult samples</td><td>Positive</td><td>0</td><td>36</td><td>36</td><td>40.0</td></tr>
<tr><td>Negative</td><td>0</td><td>54</td><td>54</td><td>60.0</td></tr>
<tr><td>Total</td><td>0</td><td>90</td><td>90</td><td></td></tr>
<tr><td>Percentage (%)</td><td>0.0</td><td>100.0</td><td></td><td></td></tr>
</table>

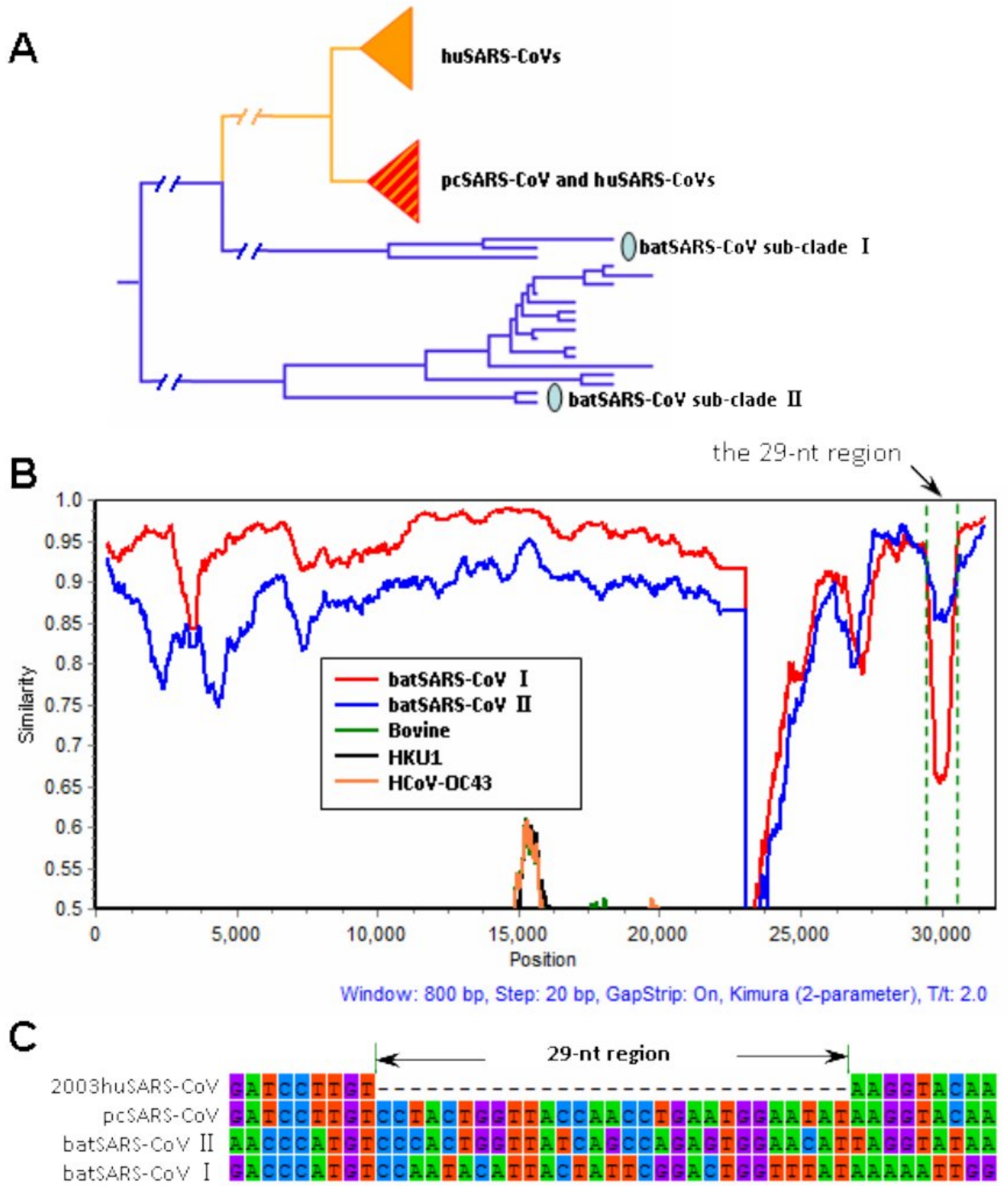

**Figure S1. Detection of recombination in the origin of human SARS-CoV. A**, the phylogeny of huSARS-CoVs, pcSARS-CoVs and batSARS-CoVs. **B**, a similarity plot analysis detected a recombination occurring in a region including the 29-nt sequence. **C**, sequence comparison of the 29-nt region between huSARS-CoV, pcSARS-CoV and batSARS-CoV. The tree of SARS-CoVs is obtained from Figure 1. The similarity plot analysis was performed using SimPlot version 3.5.

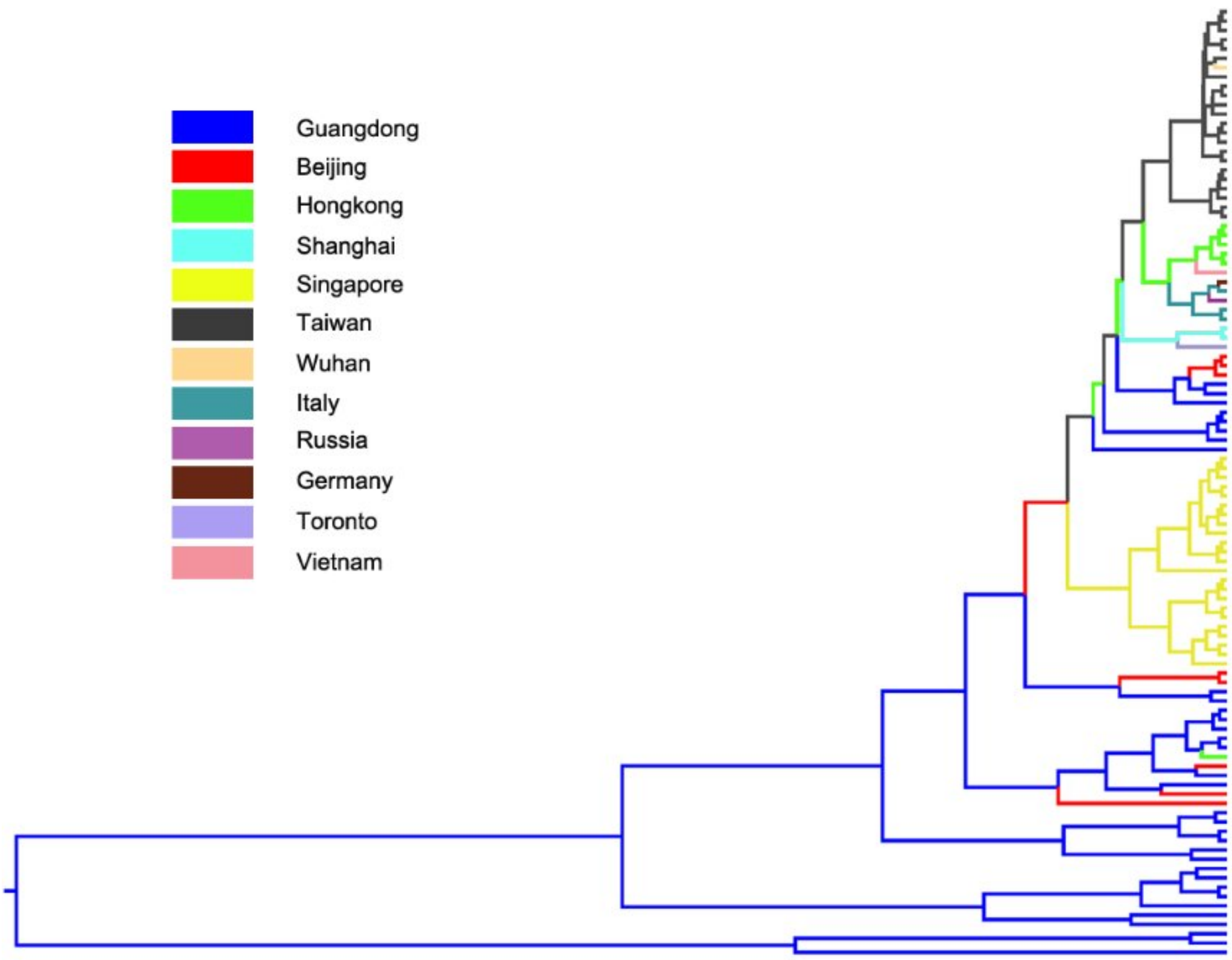

**Figure S2. MCC tree of S gene sequences from huSARS-CoV.** Ancestral geographic states were reconstructed using Bayesian phylogeographic inference framework implemented in the BEAST v1.6.2 package. The tree branches are colored according to their respective geographical locations. From the tree, we found that huSARS-CoV originated in Guangdong, and from Guangdong, SARS-CoV strains was transmitted to other provinces of China and other countries/regions. This result is consistent with the epidemiological pattern of SARS-CoV.